\documentclass[prb,twocolumn,superscriptaddress,showpacs]{revtex4}
\usepackage{graphicx}
\usepackage{dcolumn}
\usepackage{bm}

\usepackage{xcolor}
\usepackage{soul}
\usepackage{graphicx}
\usepackage{dcolumn}
\usepackage{bm}

\def\up{\uparrow}
\def\dwn{\downarrow}

\begin{document}
\title{Enhanced critical temperature, pairing fluctuation effects, and BCS-BEC crossover in a two-band Fermi gas}
\author{Hiroyuki Tajima}
\affiliation{Quantum Hadron Physics Laboratory, RIKEN Nishina Center, Wako, Saitama, 351-0198, Japan}
\author{Yuriy Yerin}
\affiliation{School of Science and Technology, Physics Division, Universit\`{a} di Camerino, 62032 Camerino (MC), Italy}
\author{Andrea Perali}
\affiliation{School of Pharmacy, Physics Unit, Universit\`{a} di Camerino, 62032 Camerino (MC), Italy}
\author{Pierbiagio Pieri}
\affiliation{School of Science and Technology, Physics Division, Universit\`{a} di Camerino, 62032 Camerino (MC), Italy}
\affiliation{INFN, Sezione di Perugia, 06123 Perugia (PG), Italy}
\date{\today}
\begin{abstract}
We study the superfluid critical temperature in a two-band attractive Fermi system with strong pairing fluctuations associated with both interband and intraband couplings. We focus specifically on a configuration where the intraband coupling is varied from weak to strong in a shallow band coupled to a weakly-interacting deeper band. The whole crossover from the Bardeen-Cooper-Schrieffer (BCS) condensation of largely overlapping Cooper pairs to the Bose-Einstein condensation (BEC) of tightly bound molecules is covered by our analysis, which is based on the extension of the  Nozi\`{e}res-Schmitt-Rink (NSR) approach to a two-band system.
In comparison with the single-band case, we find a strong enhancement of the critical temperature, a significant reduction of the preformed pair region where pseudogap effects are expected, and the entanglement of two kinds of composite bosons in the strong-coupling BEC regime. 
\end{abstract}
\maketitle
{\em Introduction} --
After the realization of the BCS-BEC crossover phenomenon in ultracold Fermi gases \cite{Regal,Bartenstein,Zwierlein}, which provided  a unified understanding of both weak coupling BCS Fermi superfluidity and Bose-Einstein condensation of molecular bosons \cite{Eagles,Leggett,Nozieres,SadeMelo,Perali2,Ohashi,Giorgini,Bloch,Strinati}, 
multi-condensates can be regarded as the next paradigmatic systems to be explored. Owing to the emergence of additional degrees of freedom of the order parameter, these multi-condensate systems can lead to a plethora of novel quantum phenomena.
While even the single-component fermionic condensate has bridged various research fields such as nuclear physics \cite{Strinati,Baldo,Stein,Lombardo,Gezerlis,mjin,Ramanan,vanWyk} and several solid-state systems \cite{Eagles,Micnas,Randeria,SadeMelo,Maly,Perali2,Tomio,Pieri,Chen,Lubashevsky,Bianconi2013,Okazaki,Kasahara,Kasahara2,Conti},
the more generic concept of multi-component BCS-BEC crossover not only builds up an interdisciplinary cross-link among strongly correlated systems but also opens a new frontier to explore the optimal configuration for high-$T_{\rm c}$ superconductivity \cite{Milosevic,Innocenti2010}. 
\par
Among the variety of unconventional superconductors recently discovered, iron-based superconducting compounds are of particular interest due to their multiband electron structure with  interband couplings producing complex order parameter symmetry and multiple energy gaps. This gives new opportunities to observe experimentally the BCS-BEC crossover in a new class of superconducting materials \cite{Lubashevsky,Bianconi2013,Okazaki,Kasahara,Kasahara2,Hashimoto,Shibauchi,Rinott}. 
Nanostructured superconductors are another promising class
of materials in which the multiband BCS-BEC crossover,
in the presence of shape resonance effects, can play a key role in the
control and enhancement of superconductivity \cite{Perali1996,Bianconi1998}.
A multi-band structure with a small Fermi surface pocket has an important role to achieve such strongly correlated crossover regime in all of these electron systems \cite{Guidini,Chubukov,Wolf,Salasnich}.
Some two-band theoretical models predict that so-called incipient bands may play an important role in several superconducting iron-based materials like FeSe intercalates and monolayers, considering them as quasi-2D systems \cite{Chen2015,Linscheid2016}. On the other hand, recent experiments indicate the presence of a 3D momentum dependence of the gap in FeSe multiband superconductors, indicating that a 3D theoretical approach may be applied for the description of the superconducting state in this compound \cite{Ge2010,Kushnirenko}.

\par
A two-band Fermi system with Josephson-like interband coupling \cite{Suhl} is also under current experimental investigation in ultracold $^{173}$Yb atomic Fermi gases near an orbital Feshbach resonance \cite{Zhang,Pagano,Hofer}.
By applying a magnetic field, the energy separation between different atom-atom scattering channels (corresponding to that between two bands in the associated model Hamiltonian) can be arbitrarily tuned, and the emergence of the BCS-BEC crossover has been theoretically predicted \cite{Iskin3,He2016,Xu,Iskin2,Zou,Mondal}. In this case, however, the situation is complicated by the presence of additional deep molecular bound states, on top of the shallow one which is responsible for the orbital Feshbach resonance~\cite{Xu,Mondal}, which make the resulting model less relevant for the physics of multiband superconductors.
\par
In this paper, we address the effects of strong pairing fluctuations on the superfluid critical temperature for a two-band system with varying intra- and interband couplings. We focus in particular on the physically relevant configuration with a shallow ``hot" band (in which the intraband coupling is varied from weak to strong) coupled with  a deep ``cold" band (with weak intraband coupling). 
For increasing hot-band coupling,  we reveal a strong amplification of the critical temperature in comparison with the single-band case, with the interband coupling assisting such amplification, but not being crucial for its occurrence. In addition, in the intermediate (crossover) region between the BCS and BEC limits, the comparison between the critical temperature and the pair-breaking temperature shows a significant shrinking of the preformed-pair region, 
implying a possible reduction of the pseudogap effects, in lines with recent experimental findings for the FeSe multiband superconductors \cite{Hanaguri}. 
Finally, in the BEC regime with finite interband coupling, an interesting coherently coupled binary mixture of composite bosons is found.
\par
\begin{figure}[t]
\begin{center}
\includegraphics[width=4.5cm]{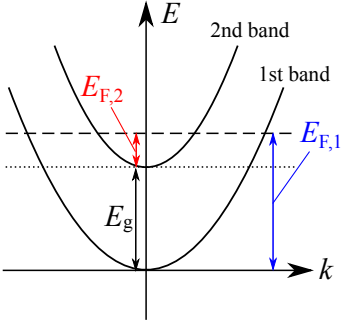}
\end{center}
\caption{The band structure considered in this work. The vertical and horizontal axes are the single-particle energy and momentum, respectively;
$E_g$ is the energy shift between the 1st ($i=1$) and the 2nd ($i=2$) band.
$E_{{\rm F},i}$ indicates the Fermi energy of $i$-band fermions in the absence of interactions.
}
\label{fig1}
\end{figure}

{\em Model} --
\label{sec2}
For the sake of generality,  we consider the following minimal model Hamiltonian~\cite{Iskin4,Iskin}  for a three-dimensional two-band Fermi system
\begin{equation}
\label{eq1}
H=\sum_{\bm{k},\sigma,i}\xi_{\bm{k},i}c_{\bm{k},\sigma,i}^{\dag}c_{\bm{k},\sigma,i} +\sum_{i,j}U_{ij}\sum_{\bm{q}}b_{\bm{q},i}^{\dag}b_{\bm{q},j}.
\end{equation}
Here, $c_{\bm{k},\sigma,i}$ is the annihilation operator of a fermion with spin $\sigma=\up,\dwn$ and band index $i=1,2$, 
$b_{\bm{q},i}^{\dag}=\sum_{\bm{k}}^{k_0}c_{\bm{k}+\bm{q}/2,\up,i}^{\dag}c_{-\bm{k}+\bm{q}/2,\dwn,i}^{\dag}$
is the pair-creation operator in the $i$-band (where $k_0$ is a momentum cutoff),  while $\xi_{\bm{k},1}=k^2/2m -\mu$, 
$\xi_{\bm{k},2}=k^2/2m +E_{\rm g}-\mu$ are the kinetic energies measured from the chemical potential $\mu$ where $E_{\rm g}$ is the energy shift between the two bands, and $m$ is the particle (effective) mass (which is assumed to be identical in the two bands).
 We also introduce the band Fermi momenta  $k_{{\rm F},i}=(3\pi^2n^0_i)^{1/3}$, defined in terms of the band densities $n^0_1$ and $n^0_2$ in the absence of any interactions and at zero temperature, with the corresponding Fermi energies $E_{{\rm F},i}=k_{{\rm F},i}^2/2m$ and temperatures $T_{{\rm F},i}$, associated with these energies. In addition to that we will use the total Fermi momentum $k_{\rm F,t}=(3\pi^2n)^{1/3}$  defined by the total number density $n$, which is kept fixed. 
We set $k_{\rm B},\hbar$, and the system volume $V$ equal to one. Fig.~\ref{fig1} summarizes graphically our two-band configuration.

We express the intraband coupling $U_{ii} (<0)$ in terms of the intraband scattering length $a_{ii}$ \cite{Iskin4,Iskin}:
\begin{equation}
\label{eq2}
 \frac{m}{4\pi a_{ii}}=\frac{1}{U_{ii}}+\sum_{\bm{k}}^{k_0}\frac{m}{k^2}.
\end{equation}
The momentum cutoff $k_0$ is  considered to be much larger than the average interparticle distance, corresponding to a short-range condition on the interaction.  Specifically, we take $k_0= 100 k_{\rm F,t}$.
We choose the relatively large energy shift $E_{\rm g}=0.75E_{\rm F,1}=3E_{\rm F,2}$, which implies
 $k_{\rm F,1}=\left(8/9\right)^{1/ 3}k_{\rm F,t}$ and $k_{\rm F,2}=\left(1/9\right)^{1/ 3}k_{\rm F,t}$.
We focus on the situation in which the shallow hot band ($i=2$) undergoes the BCS-BEC crossover, whereas the intraband coupling in the deep cold band ($i=1$) remains weak. In our configuration, this is achieved by taking $U_{22}= 1.1 U_{11}$, which gives $(k_{\rm F,1}a_{11})^{-1}$ ranging between $\simeq-8$ and $\simeq-4$ when we change $(k_{\rm F,2}a_{22})^{-1}$ from the weak coupling (BCS) to the strong-coupling (BEC) regime. We recall that for a single-band system the BCS-BEC crossover is driven by the dimensionless parameter $(k_{\rm F} a)^{-1}$, which ranges  from $(k_{\rm F} a)^{-1}\lesssim -1$ in the weak-coupling (BCS) regime to $(k_{\rm F} a)^{-1}\gtrsim 1$ in the strong-coupling (BEC) regime.

The interband couplings are equal and real (guaranteeing the hermiticity of the Hamiltonian). For convenience, we introduce the dimensionless interband coupling $\tilde{U}_{12}$  by setting $U_{12}= \tilde{U}_{12} (k_{\rm F,t}/k_0)^2 E_{\rm F, t}/n$, with $E_{\rm F,t}=k_{\rm F,t}^2/2m$. In this way, when $\tilde{U}_{12}$ ranges from 0 to 5,  the effects of the interband coupling on the quantities of interest will turn to vary from weak to strong.

\begin{figure}[t]
\begin{center}
\includegraphics[width=6cm]{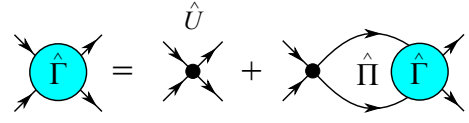}
\end{center}
\caption{Diagrammatic representation of the many-body $T$-matrix $\hat{\Gamma}$. $\hat{U}$ and $\hat{\Pi}$ are the $2\times2$ matrices of coupling constants and pair susceptibilities, respectively.
}
\label{fig2}
\end{figure}

\par
{\em Formalism and results} --
The NSR formalism has been widely used for studying the BCS-BEC crossover in a single-band system.
Ref.~\onlinecite{Iskin} generalized the formalism to the two-band case, but only at the formal level. Here, we present an explicit numerical solution of the associated equations, and  study the effect of pairing fluctuation at finite temperature in a two-band system across the whole BCS-BEC crossover. 

Let us first briefly summarize the main equations of the NSR formalism for two-band systems.
The sum of ladder diagrams defines the many-body $T$-matrix  $\hat{\Gamma}$, as  represented in  Fig.~\ref{fig2}, which satisfies
\begin{eqnarray}
\label{eq3}
\hat{\Gamma}(\bm{q},i\nu_l)=[1+\hat{U}\hat{\Pi}(\bm{q},i\nu_l)]^{-1}\hat{U},
\end{eqnarray}
where $\hat{U}$ is the $2 \times 2$ matrix constructed with the interaction parameters $U_{ij}$, and 
\begin{eqnarray} 
\label{eq5}
\hat{\Pi}(\bm{q},i\nu_l)=
\left(\begin{array}{cc}
\Pi_{11}(\bm{q},i\nu_l) & 0 \\
0 & \Pi_{22}(\bm{q},i\nu_l) 
\end{array}
\right),
\end{eqnarray}
where $\nu_l=2l\pi T$ ($l$ integer) is a boson Matsubara frequency at temperature $T$ and 
\begin{eqnarray}
\label{eq6}
\Pi_{ii}(\bm{q},i\nu_l)=T\sum_{\bm{k},i\omega_s}^{k_0}G^{0}_{i}(\bm{q}-\bm{k},i\nu_l-i\omega_s)
G^{0}_{i}(\bm{k},i\omega_s).
\end{eqnarray}
Here $G^{0}_{i}(\bm{k},i\omega_s)=1/(i\omega_s-\xi_{\bm{k},i})$ is the bare Green's function of a $i$-band particle and $\omega_s=(2s+1)\pi T$ ($s$ integer) is a fermion Matsubara frequency.
\par
The critical temperature $T_{\rm c}$ is determined by the  Thouless criterion \cite{Thouless},  namely the divergence of $\hat{\Gamma}(0,0)$, corresponding to the condition
\begin{eqnarray}
\label{eq7}
{\rm det}\left[1+\hat{U}\hat{\Pi}(\bm{q}=0,i\nu_l=0)\right]=0.
\end{eqnarray}
This equation needs to be solved together with the particle number equation $n=-\partial\Omega/\partial\mu$,  where $\Omega$ is obtained by adding the thermodynamic potential constructed from the ladder diagrams to the free one~\cite{Nozieres,Iskin}.  
It is easy to show that in the present two-band case, -$\partial\Omega/\partial\mu$ can be expressed as the sum $n_1+n_2$ of the densities in the two bands, with 
\begin{equation}
\label{eq9}
n_{i}=2 \left[\sum_{\bm{k}}f(\xi_{\bm{k},i})+T\sum_{\bm{k},{i\omega_s}}^{k_0}G^{0}_{i}(\bm{k},i\omega_s)^2\Sigma_i(\bm{k},i\omega_s)\right],    
\end{equation}
where $f$ is the Fermi function at temperature $T$ and we have introduced the self-energy 
\begin{equation}
\Sigma_i(\bm{k},i\omega_s)=T\sum_{\bm{q},i\nu_l} \Gamma_{ii}(\bm{q},i\nu_l)G^{0}_{i}(\bm{q}-\bm{k},i\nu_l-i\omega_s)
\end{equation}
for fermions in the band $i$.
At fixed $n$, $T$, and interaction parameters, the inversion of the number equation $n=n_1(\mu)+n_2(\mu)$ determines the chemical potential $\mu$.
\par
\begin{figure}[t]
\begin{center}
\includegraphics[width=6cm]{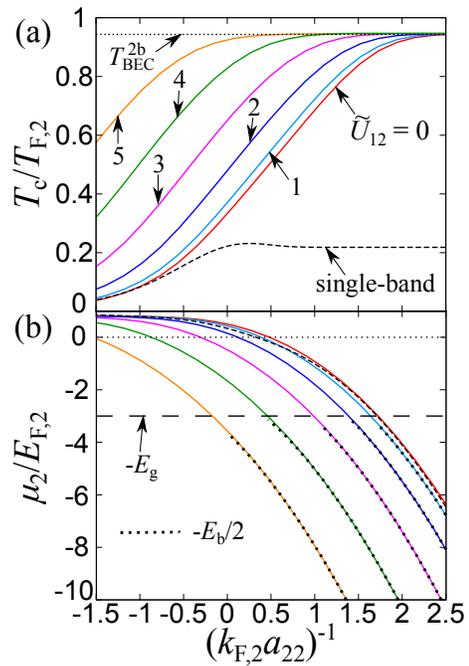}
\end{center}
\caption{(a) Critical temperature $T_{\rm c}$ and (b) chemical potential $\mu_2\equiv\mu-E_{\rm g}$ at $T_c$ vs.~$(k_{\rm F,2}a_{22})^{-1}$ at different interband couplings  $\tilde{U}_{12}$ (we use the same line-styles in both panels).
The constant lines  $T_{\rm BEC}^{\rm 2b}=0.943T_{\rm F,2}$ and  $-E_{\rm g}$, as well as the curves $-E_{\rm b}/2$, with the two-body binding energy $E_{\rm b}$ obtained from Eq.~(\ref{eq12}) at different $\tilde{U}_{12}$ are also reported for reference.
}
\label{fig3}
\end{figure}
Figure \ref{fig3} shows the overall critical temperature $T_{\rm c}$ and the corresponding chemical potential  $\mu-E_{\rm g}\equiv\mu_2$  measured from the bottom of the hot band as a function of $(k_{\rm F,2}a_{22})^{-1}$.   We compare the results of the two-band system with the case in which the hot band is considered as a single band, with density fixed to $n^0_2$ (for our choice of $E_{{\rm F},i}$, $n^0_2=n/9$). Note that the single-band case differs from simply setting $U_{12}=0$ in the two-band system, for which a particle transfer between the two bands is possible.  
One sees indeed that for vanishing interband interaction, while in the weak-coupling (BCS) regime 
$T_{\rm c}$ and $\mu_{2}$ essentially coincide with the corresponding single-band results, in the intermediate and strong coupling regions,  $T_{\rm c}$ is greatly enhanced in comparison with the single-band case. 
In the two-band system, the pairing attraction in the hot band drains particles from the cold band to lower the overall free energy.  The critical temperature is then enhanced, until 
in the strong-coupling limit the asymptotic value $T_{\rm BEC}^{\rm 2b}=0.218\left(k_{\rm F,t}/k_{\rm F,2}\right)^2T_{\rm F,2}=0.943T_{\rm F,2}$ is reached,
corresponding to the condensation temperature for a gas of non-interacting bosons of density $n/2$ and mass $2m$.
In this limit, $\mu_2$ coincides with $- E_{\rm b}/2$, where $E_{\rm b}$ is the two-body binding energy, which for the decoupled system is given by $E_{\rm b}=(ma_{22}^2)^{-1}$ . 
\par
Even more interesting is the situation with a finite interband coupling. In this case the BEC limit can be effectively reached  even for rather weak values of the intraband coupling $(k_{\rm F,2}a_{22})^{-1}$. This behavior can be understood by  noting that the Thouless criterion (\ref{eq7})  can be rewritten as $1+U_{{\rm eff}}\Pi_{22}(0,0)=0$,  where
\begin{equation}
\label{eq11}
U_{\rm eff}=U_{22}-U_{12}^2\Pi_{11}(0,0)/[1+U_{11}\Pi_{11}(0,0)],
\end{equation}
is the effective interaction determining $T_c$. 
It is the second term in Eq.~(\ref{eq11}) which leads to a significant increase of the effective pairing interaction  when $U_{12}$ increases. It can be shown in particular 
that in our configuration this term effectively corresponds to a shift  $\simeq -0.5 k_{\rm F, 1} a_{11} \tilde{U}_{12}^2$ of the dimensionless coupling $(k_{\rm F,2}a_{22})^{-1}$ toward stronger couplings.  
\begin{figure}[t]
\begin{center}
\includegraphics[width=6cm]{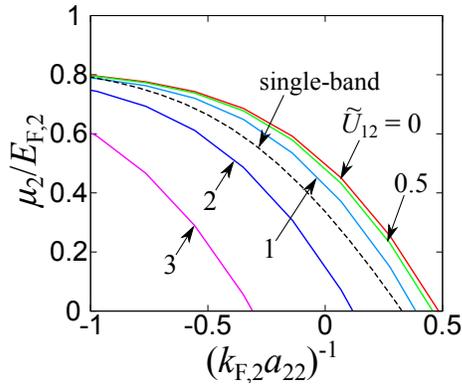}
\end{center}
\caption{Chemical potential $\mu_{2}\equiv \mu-E_{\rm g}$  vs.~$(k_{\rm F,2}a_{22})^{-1}$ at $T=T_{\rm c}$ near the unitarity limit $(k_{\rm F,2}a_{22})^{-1}=0$.
}
\label{fig4}
\end{figure}
The effect of the interband coupling can also be seen at the two-body level on the binding energy $E_{\rm b}$, which for the coupled system can be obtained from the equation determining the pole of the two-body $T$-operator on the real  energy axis
\begin{equation}
\label{eq12}
\left(1+U_{11}\Pi_{11}^0\right)\left(1+U_{22}\Pi_{22}^0\right)-U_{12}^2\Pi_{11}^0\Pi_{22}^0=0,
\end{equation}
where the vacuum particle-particle bubbles $\Pi_{ii}^0$ are obtained from $\Pi_{ii}(\bm{q},i\nu_l)$ by setting  $\bm{q}=0,i\nu_l\rightarrow -(E_{\rm b}-2 E_{\rm g})-2\mu$ and then taking the vacuum limit $\mu/T\rightarrow -\infty$.  One then gets from Eq.~(\ref{eq6})
\begin{equation}
\label{eq13}
\Pi_{ii}^0=\frac{mk_0}{2\pi^2}\left[1-\frac{\sqrt{m|E_i|}}{k_0}{\rm arctan}\left(\frac{k_0}{\sqrt{m|E_i|}}\right)\right],
\end{equation}
where we have defined  $E_{i}\equiv-E_{\rm b}+2E_{\rm g}\delta_{i,1}$. Note that the binding energy $E_{\rm b}$ is referred to the bottom of the upper band, and $E_{\rm b} > 2 E_{\rm g}$ for $U_{12}\neq 0$.
One sees in Fig.~\ref{fig3}(b) that when $\tilde{U}_{12}$ increases, the binding energy $E_{\rm b}$ also increases, and $\mu_2$ approaches the BEC limit $-E_{\rm b}/2$ at progressively weaker intraband couplings $(k_{\rm F,2}a_{22})^{-1}$.
\par
Figure \ref{fig4} focuses on the behavior of $\mu_{2}$ near the unitarity limit $(k_{\rm F,2}a_{22})^{-1}=0$. 
One can see that in the absence or for weak interband coupling, the chemical potential for the two-band  system is larger than for a single-band, such that the crossing of the bottom of the upper band is shifted to larger values of  $(k_{\rm F,2}a_{22})^{-1}=0$.
Physically, this is explained by the Pauli-blocking effect due to the occupied states in the cold band which acts to retard the ``bosonization" of the Cooper pairs. However, when the interband coupling increases, it overcomes this quantum statistical effect and eventually shifts the bosonization to smaller intraband coupling values.
\par 
\begin{figure}[t]
\begin{center}
\includegraphics[width=6cm]{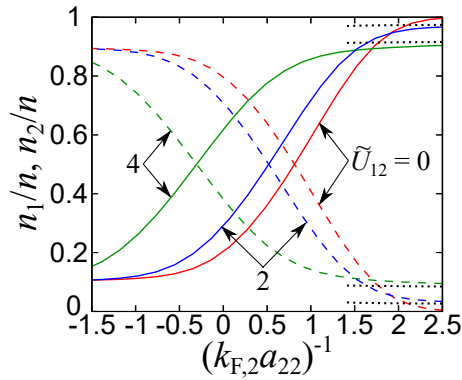}
\end{center}
\caption{Number densities $n_i$  (in units of $n$)  in the lower ($i=1$, dashed line) and upper ($i=2$, solid line)  bands at $T_c$ vs.~$(k_{{\rm F},2}a_{22})^{-1}$  for different values of $\tilde{U}_{12}$.
The dotted lines correspond to the ratios $||\psi_i||^2/(||\psi_1||^2+||\psi_2||^2)$ in the two-body problem (see text) for $i=1$, $\tilde{U}_{12}=2,4$ and $i=2, \tilde{U}_{12}=4,2$, from bottom to top.
}
\label{fig5}
\end{figure}
Figure~\ref{fig5} shows how the fermions distribute  between the two bands when the coupling  $(k_{\rm F,2}a_{22})^{-1}$ is varied. 
For different values of the interband coupling $\tilde{U}_{12}$, one observes in general a progressive transfer of particles from the cold band to the hot one as $(k_{\rm F,2}a_{22})^{-1}$ increases. However, while in the BEC limit one has a full  transfer for $\tilde{U}_{12}=0$,  at finite interband coupling the transfer is only partial. This behavior can be understood by solving the two-body bound-state equation $H_{\rm rel} \Psi = E \Psi $ for the relative motion wave-function  $\Psi(r)=\left(\psi_{1}(r), \psi_{2}(r)\right)$, where $\psi_1$ and $\psi_2$ are the components in the two bands, $H_{\rm rel}$ is the two-body relative motion Hamiltonian associated with the many-body Hamiltonian (\ref{eq1}), and $r$ is the relative distance.  In this way one finds that when $\tilde{U}_{12}\neq 0$ the bound-state solution has components in both bands, and the asymptotic value of $n_i/n$ in the BEC limit of the many-body problem coincides with the ratio $||\psi_i||^2/(||\psi_1||^2+||\psi_2||^2)$  in the two-body problem (dotted lines in Fig.~\ref{fig5}). Here, $||\psi_i||^2=\int \! d^3 r |\psi_i(r)|^2$; note that the condition $E_{\rm b}>2E_{\rm g}$ is required to have $||\psi_1|| < \infty$.      

In this extreme BEC regime, the bosons condensing at $T_c$ are a coherent superposition of two molecular states in the two bands, with  wave functions $\psi_{i}(r)\propto e^{-\sqrt{m|E_{i}|}r}/r$ and  sizes $R_{1}=1/\sqrt{m|2E_{\rm g}-E_{\rm b}|}$ and $R_{2}=1/\sqrt{m|E_{\rm b}|}$. In the less extreme regime whereby the chemical potential $\mu_2$ is 	between the bottom of the two bands, the two-body bound state is effectively replaced  by a quasi-bound state, i.e., a resonance whose energy $E_{\rm r}$  essentially determines the value of the chemical potential ($\mu_2 \simeq  E_{\rm r}/2$). The energy $E_{\rm r}$ corresponds to a peak (narrow for small $\tilde{U}_{12}$) of the two-body $T$-matrix,  while only the solution of the many-body problem yields the relative distribution of particles between the two bands. 
Interestingly,  unusual vortex configurations with non-triangular geometry, stripes, or multi-quantum-vortex lattices are expected to occur in two-band superconductors with non-equal pair sizes \cite{Babaev2005,Wolf} and two-species BEC \cite{Mueller2002,Kuopanportti2012}. 
\begin{figure}[t]
\begin{center}
\includegraphics[width=6cm]{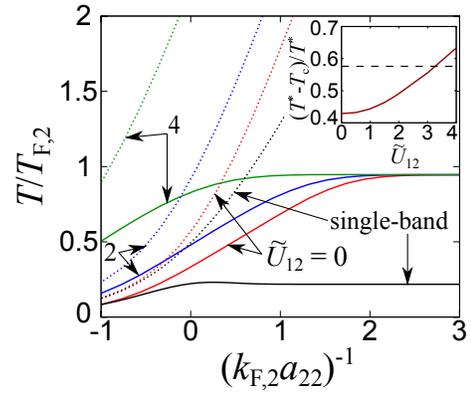}
\end{center}
\caption{Critical temperature $T_c$ (solid lines) and pair-breaking temperature $T^*$ (dotted lines) as estimated by the mean-field critical temperature,  vs.~$(k_{\rm F,2}a_{22})^{-1}$ for different values of $\tilde{U}_{12}$. The inset shows the ratio $(T^*-T_c)/T^*$ near unitarity [$(k_{\rm F,2}a_{22})^{-1}= 0.07$] as a function of $\tilde{U}_{12}$. The corresponding value for the single-band case is also reported (dashed line).} 
\label{fig6}
\end{figure}
\par
In Fig.~\ref{fig6} the preformed pair region of the phase-diagram between the
mean field temperature $T^*$ and the superfluid critical temperature
$T_c$ is investigated for different interband coupling $\tilde{U}_{12}$.
This region is of particular interest because 
below $T^*$, pseudogap phenomena and molecular-like pairing
are expected to appear, with detectable signatures in the single-particle excitation
spectra, depending on the intraband coupling strength \cite{NP2010,PRL11}.
Also, the amount of pair fluctuations in this region is responsible
for the detrimental suppression of the superfluid critical temperature.
The pairing temperature $T^*$ is strongly enhanced by increasing
$\tilde{U}_{12}$. It is already larger than the corresponding temperature
for the single-band case for vanishing interband coupling, because of the larger
chemical potential, see Fig.~\ref{fig4}. On the other hand, for
intraband couplings in the hot band close to unitarity and in a sizable
range of $\tilde{U}_{12}\lesssim 3$, the preformed pair region is reduced with respect to the single-band case,
as quantified by the temperature window $(T^*-T_c)/T^*$, reported in the inset of Fig. ~\ref{fig6}. This can be
connected with the recent  experiment in the multiband FeSe superconductor where BCS-BEC crossover signatures have been confirmed while a pseudogap was not detected~\cite{Hanaguri}.
\acknowledgments
H.~T. thanks C.~A.~R.~S\'{a} de Melo for useful discussion during the International Conference on Multi-Condensate Superconductivity and Superfluidity in Solids and Ultra-cold Gases (Multisuper 2018) and acknowledges the hospitality of the Physics Division at the U.~of Camerino, where most of this work was done.
H.~T. also thanks Y. Kondo for discussions. 
H.~T. was supported by a Grant-in-Aid for JSPS fellows (No.17J03975).
This work was supported by the Italian MIUR through the PRIN 2015 program (Contract No. 2015C5SEJJ001) and by the RIKEN iTHEMS program.

\end{document}